\documentclass{mn2e}
\usepackage{times,graphicx,astrobib,amssymb,aas_macros}

\newcommand{\be}{\begin{equation}}
\newcommand{\e}{\end{equation}}
\newcommand{\bear}{\begin{eqnarray}}
\newcommand{\ear}{\end{eqnarray}}

\newcommand{\f}{\frac}
\newcommand{\de}{{\rm d}}

\newcommand{\matrixsymbol}{\sf}

\begin{document}

\title[Joint quasar-CMB constraints on reionization history]
{Joint QSO-CMB constraints on reionization history}
\author[Mitra, Choudhury \& Ferrara]
{Sourav Mitra$^1$\thanks{E-mail: smitra@hri.res.in},~
T. Roy Choudhury$^1$\thanks{E-mail: tirth@hri.res.in}~
and
Andrea Ferrara$^2$\thanks{E-mail: andrea.ferrara@sns.it}
\\
$^1$Harish-Chandra Research Institute, Chhatnag Road, Jhusi, Allahabad 211019, India\\
$^2$Scuola Normale Superiore, Piazza dei Cavalieri 7, 56126 Pisa, Italy
}

\maketitle

\date{\today}

\begin{abstract}
Based on the work by \citeN{2010arXiv1011.2213M}, we obtain 
model-independent constraints on 
reionization from cosmic microwave background (CMB) and QSO absorption line data by 
decomposing the function $N_{\rm ion}(z)$ (the 
number of photons entering the IGM per baryon in collapsed objects) into its 
principal components. 
The main addition in this work is that for the CMB data set, we explicitly include the 
angular power spectra $C_l$ for 
TT, TE and EE modes in our analysis which seem to contain somewhat 
more information than taking the electron scattering
optical depth $\tau_{\rm el}$ as a single data point.
Using Markov Chain Monte Carlo methods, we find that all the quantities related to 
reionization can be severely constrained at $z<6$ whereas a broad range of 
reionization histories at $z > 6$ are still permitted by the current data sets. 
With currently available data from WMAP7, we constrain 
$0.080 < \tau_{\rm el} < 0.112$ (95\% CL) and also
conclude that reionization is 50\% complete between 
$9.0 < z(Q_{\rm HII} = 0.5) < 11.8$ (95\% CL) and is 
99\% complete between $5.8 < z(Q_{\rm HII} = 0.99) < 10.4$ (95\% CL).
With the forthcoming 
PLANCK data on large-scale polarization (ignoring effect of foregrounds), 
the $z > 6$ constraints will be improved considerably, e.g., 
the $2-\sigma$ error on $\tau_{\rm el}$ will be reduced 
to 0.009 and the uncertainties on $z(Q_{\rm HII} = 0.5)$ and 
$z(Q_{\rm HII} = 0.99)$ would be $\sim 1$ and 3 (95\% CL), respectively. 
For more stringent constraints on reionization at $z > 6$, one has to rely on data
sets other than CMB. Our method will be useful in such case 
since it can be used for non-parametric 
reconstruction of reionization history with arbitrary data sets.

\end{abstract}


\section{Introduction} 

In the past few years, the understanding of reionization process has become increasingly  
sophisticated in both the observational and theoretical communities (for reviews, see, 
\citeNP{2001ARA&A..39...19L,2001PhR...349..125B,2006astro.ph..3149C,2009arXiv0904.4596C,2006PhR...433..181F,2006ARA&A..44..415F}  
and the references therein), thanks to the availability of good quality data related to reionization. 
Mainly, the observations by the {\it Wilkinson Microwave Anisotropy Probe} (WMAP) satellite 
of cosmic microwave background (CMB) and highest redshift QSOs put very tight constraints 
on the reionization history of the universe. The WMAP seven-year observation manifests the 
Thomson scattering optical depth $\tau_{\rm el}=0.088\pm0.015$ \cite{arXiv:1001.4635v2} with 
the simple assumption that the universe was reionized instantaneously. However, recent 
studies suggest that reionization process is too complex to be described as a sudden process. 
In fact, the physical processes relevant to reionization are 
so complex that neither the analytical nor the numerical simulations alone can capture 
the overall picture. That's why, it is often studied using semi-analytical models of 
reionization, with limited computational resources. 

The major uncertainty in modeling any semi-analytical reionization scenario is to model 
the parameter $N_{\rm ion}$, the number of photons entering the IGM per baryon in collapsed 
objects, which can be a function of redshift $z$. In analytical studies, $N_{\rm ion}(z)$ 
is either taken to be a piecewise constant function \cite{2003ApJ...586..693W,2005MNRAS.361..577C}, 
parameterized using some known functions \cite{2003ApJ...599..759C,2010MNRAS.408...57P}, 
modeled using a physically-motivated prescription \cite{2006MNRAS.371L..55C}, or taken 
to be an arbitrary function of $z$ and decomposed into its principal components using the 
principal component analysis (\citeNP{2010arXiv1011.2213M}, hereafter Paper I). 

The principal component method 
has been applied to study the constraints on reionization from large-scale CMB polarization 
(\citeNP{2008ApJ...672..737M,2003PhRvD..68b3001H}). It is well established that, the 
inhomogeneity signature of reionization is expected to contribute to the CMB temperature 
and polarization anisotropies (\citeNP{2000ApJ...529...12H,2005MNRAS.360.1063S,2006NewAR..50..909I,2007ApJ...657....1M}). 
In fact, the CMB power spectra contain more information than the optical depth 
integrated over the whole ionization history \cite{2003PhRvD..68b3001H}. So it is 
worth asking what we can ultimately expect to learn about the reionization model with 
principal component technique from the current CMB data sets instead of single optical depth 
data. 

In our previous work, we made a preliminary attempt to constrain $N_{\rm ion}(z)$ using 
PCA and estimated the uncertainties in the reionization history.
The main difference of our work with other PCA of reionization history using CMB data 
\cite{2008ApJ...672..737M,2003PhRvD..68b3001H} is that we use a self-consistent model of 
reionization and include data sets other than CMB (e.g., QSO absorption lines) in the analysis. 
Such an analysis should give us a handle in not only constraining the
evolution of the electron fraction $x_e(z)$ (as is done in usual reionization related studies
using CMB data) but also in constraining 
the evolution of source properties like galactic IMF, star-formation history, and escape fraction 
of ionizing photons.

In Paper I, we found that to model 
$N_{\rm ion}(z)$ over the range $2 < z < 14$ one should include the first 5 principal components 
with smaller uncertainties. We concluded that a wide range of reionization scenarios 
are allowed by the data sets of photoionization rates, redshift distribution of Lyman-limit 
systems and the electron scattering optical depth from WMAP7. 
In this paper, we extend our previous work to study the effect of inclusion of the 
angular power spectra $C_l$ of the CMB temperature (T) and polarization (E)
modes. Using the available WMAP7 data on $C_l^{\rm TT, TE, EE}$, we  study the
present constraints on reionization history. We also forecast the errors on
reionization history as would be determined by future observations
of large scale polarization signal by PLANCK\footnote{http://www.esa.int/SPECIALS/Planck/index.html}.

The paper is organized as follows. In Section \ref{sec:modelandmethod}, we discuss about 
the features of the semi-analytical model of reionization and its modifications for including 
CMB data. We also outline the basic theory of the principal component analysis in this section. 
We describe our results of the principal 
component approach to reionization model with large-scale \textit{E}-mode data in 
Section \ref{sec:results}. In this Section, using the Markov Chain Monte Carlo methods, 
we examine our model for both 7-year WMAP data and simulated PLANCK data. Finally we 
summarize our main findings and conclude in Section \ref{sec:summary}. 

\section{Model and method of statistical analysis}
\label{sec:modelandmethod}

\subsection{Semi-analytical model of reionization with PCA}
\label{sec:pcamodel}

We first describe the method used in our previous work (Paper I)
which was based on the semi-analytical model of reionization developed in
\citeN{2006MNRAS.371L..55C} and \citeN{2005MNRAS.361..577C}. The main features
of the model are:

\begin{itemize}

\item The model follows the ionization and thermal histories of neutral, HII and HeIII 
regions simultaneously and self-consistently taking the IGM inhomogeneities by adopting 
a lognormal distribution according to the method outlined in \citeN{2000ApJ...530....1M}. 

\item Given the collapsed fraction $f_{\rm coll}$ of dark matter haloes, 
this model calculates the production rate of ionizing photons in the IGM as
\be
\dot{n}_{\rm ph}(z) = n_b
N_{\rm ion} \f{\de f_{\rm coll}}{\de t}
\e
where $n_b$ is the total baryonic number density in the IGM, $N_{\rm ion}$ 
is the number of photons entering the IGM per baryon in collapsed objects. 
The parameter $N_{\rm ion}$ can actually
be written as a 
combination of various other parameters which characterize
the star-forming efficiency 
(fraction of baryons within
collapsed haloes going into stars), the fraction of photons
escaping into the IGM,  and the number of photons emitted per  
frequency range per unit mass of stars (which depends on the stellar IMF and
the corresponding stellar spectrum).

\item The model computes \textit{radiative feedback} (suppressing star 
formation in low-mass haloes using a Jeans mass prescription) 
self-consistently from the evolution of the thermal properties of the IGM. 
The corresponding filtering scale, which depends on the temperature
evolution of the IGM, is found to be typically around $\sim 30$ km s$^{-1}$.
We should mention here that minimum mass of star-forming haloes is 
much larger in ionized regions than in the neutral regions because of 
this radiative feedback. For that, this model takes the filter mass for 
ionized region and atomic cooling (i.e. small halo) for the neutral region.

\item In Paper I, we assume $N_{\rm ion}$ to be an unknown 
function of $z$ and decompose it 
into principal components. These principal components essentially filter out 
components of the model which are most sensitive to the data and thus they are 
the ones which can be constrained most accurately. 
We carry out our analysis assuming that 
only one population of stars contribute to the
ionizing radiation; any change in the characteristics of these stars 
over time or the chemical feedback prescription would be accounted for indirectly by 
the evolution of $N_{\rm ion}$. 
We also include the contribution of quasars 
at $z < 6$ assuming that they have negligible effects on IGM at higher redshifts, 
but are significant sources of photons at $z\lesssim 4$.

\item Usually, the model is constrained by comparing with a variety
of observational data, but to keep the analysis simple, we used the three 
main data sets in our earlier work, namely,  
the photoionization rates $\Gamma_{\rm PI}$ obtained using 
    Ly$\alpha$ forest Gunn-Peterson optical depth observations and a large 
    set of hydrodynamical simulations \cite{2007MNRAS.382..325B}, 
    the redshift distribution of LLS $\de N_{\rm LL}/\de z$ at $z \sim 3.5$ 
    \cite{2010ApJ...718..392P} and the WMAP7 data on electron scattering optical depth $\tau_{\rm el}$
    \cite{arXiv:1001.4635v2}. It should be mentioned that in this work, 
    we have used the data on $C_l$'s rather than the constraints on $\tau_{\rm el}$, 
    which will be described in the next subsection.
    
    \item The free parameters used in the model are the coefficients 
    related to the principal components of $N_{\rm ion}$ and $\lambda_0$ (the normalization which 
determines the mean free path of photons). The constraints on $N_{\rm ion}$ were
obtained by marginalizing over $\lambda_0$. The cosmological parameters were taken to
be fixed (given by the best-fit WMAP7 values) and not varied at all.

\end{itemize}

\subsection{Data sets and free parameters}
\label{sec:datasets}

The major modifications made in this work compared to our previous one are
related to how we treat the CMB data sets.
Note that in Paper I, $\tau_{\rm el}$ constraint was treated as a 
single data point which 
can be thought as a simplification of the CMB polarization observations at low 
multipole moments \cite{2008MNRAS.385..404B}. 
We know that, the amplitude of fluctuations in the large-scale (low-$l$) \textit{E}-mode component 
of CMB polarization provides the current best constraint on $\tau_{\rm el}$. Using the data 
from seven year WMAP and the assumption of instantaneous reionization, 
\citeN{arXiv:1001.4635v2} find $\tau_{\rm el}=0.088\pm0.015$. However, recent theoretical and 
numerical studies suggest that reionization is a fairly complex process. In that case, the 
low-$l$ \textit{E}-mode spectrum depends not just on $\tau_{\rm el}$ but also on the detailed redshift 
evolution of the number density of free electrons in the IGM, $x_e(z)$. For fixed values of 
$\tau_{\rm el}$ and all other relevant cosmological parameters, differences in $x_e(z)$ can 
affect the shape of the large-scale \textit{E}-mode angular power spectrum up to multipoles $l\simeq40-50$. 
Because of this dependence, measurements of the low-$l$ $C_l^{\rm EE}$ should place at least 
weak constraints on the overall reionization history in addition to the constraint on the total 
optical depth.

Now, in our model, the change in the parameter $N_{\rm ion}(z)$ directly corresponds to the 
change in $x_e(z)$ i.e. in other words, changes in $N_{\rm ion}$ can affect the shape of low-$l$ 
$C_l^{\rm EE}$. So, incorporating the data sets for large-scale EE polarization signal in 
our model can provide important information about the evolution of $N_{\rm ion}$ at $z>6$ beyond 
the information about $\tau_{\rm el}$. Our hope is this may be most useful for distinguishing 
the models of reionization with different ionization histories but same optical depth. 
Keeping this in mind, it
would be more prudent to work with the actual data related to the
angular power spectra $C_l$ and obtain constraints on reionization parameters; 
the constraint on $\tau_{\rm el}$ will be determined a posteriori.

The moment we include the $C_l$'s (TT+TE+EE) in our analysis, we realize that 
parameters related to reionization may have strong degeneracies with (some of)
the cosmological parameters and hence 
constraints on reionization without varying cosmological parameters would be
misleading. On the other hand, including all the cosmological parameters
in the analysis would increase the number of free parameters to a large
number. Usually, it is found that $\tau_{\rm el}$ is strongly
degenerate with the normalization of the matter power spectrum
$\sigma_8$ and also with the slope $n_s$ \cite{2003ApJS..148..175S}. Hence, it may be
worthwhile to verify whether we can carry out our analysis by varying 
only these two parameters
(in addition to the parameters related to reionization model) and keeping 
all the other cosmological parameters fixed to their
mean value.

To verify the viability of this method, we re-do the analysis of WMAP7 data with 
instantaneous
reionization history (as in \citeNP{arXiv:1001.4635v2}.
We assume the universe to be described by 
a flat cold dark matter model with a cosmological 
constant ($\Lambda$CDM) which is parametrized by six
parameters ($\Omega_b h^2, \Omega_{\rm DM} h^2, H_0, n_s, \sigma_8, \tau_{\rm el}$).
We then carry out the standard MCMC analysis \cite{2003ApJS..148..195V} 
first varying all six parameters and then
keeping all but $\sigma_8, n_s$ and $\tau_{\rm el}$ fixed to their best-fit values. 
The results are shown in Table \ref{tab:wmap_likelihood}.
It is clear that though the uncertainties on 
$n_s$ and $\sigma_8$ are reduced considerably because 
of not varying the other three parameters, 
the constraints on $\tau_{\rm el}$ are relatively unchanged. There
is only a slight ($\lesssim 15$ percent) decrease in the error-bars, thus indicating
that the parameters related to reionization are
only moderately degenerate with the other cosmological parameters. 
Hence, we can carry our analysis with the other cosmological parameters fixed
keeping in mind that the uncertainties in reionization history would possibly be slightly
underestimated. This approach is similar to what is adopted by
\citeN{2007ApJ...657....1M}.

\begin{table}
\begin{tabular}{ccc}
Parameters & \multicolumn{2}{c}{Mean value and $1-\sigma$ errors}\\
& varying all 6 parameters & varying only 3 parameters\\
\hline
$\Omega_b h^2 \times 10^2$ & $2.249^{+0.056}_{-0.057}$ & $2.249$ (fixed) \\
$\Omega_{\rm DM} h^2$ & $0.1120^{+0.0056}_{-0.0056}$ & $0.1120$ (fixed)\\
$H_0$ & $70.4^{+2.5}_{-2.5}$ & $70.4$ (fixed) \\
$n_s$ & $0.967^{+0.014}_{-0.014}$ & $0.969^{+0.007}_{-0.007}$ \\
$\sigma_8$ & $0.811^{+0.030}_{-0.031}$ & $0.816^{+0.013}_{-0.013}$\\
$\tau_{\rm el}$ & $0.088^{+0.007}_{-0.008}$ & $0.088^{+0.006}_{-0.007}$\\
\hline
\end{tabular}
\caption{Mean value of parameters and the corresponding errors for
a flat $\Lambda$CDM 
cosmological model with instantaneous reionization. The results are shown
when all six parameters are varied and when all but 
$n_s, \sigma_8$ and $\tau_{\rm el}$ are kept fixed to their
mean values.
}
\label{tab:wmap_likelihood}
\end{table}

In addition to the CMB data, we have also 
included the more recent measurements of $\de N_{\rm LL}/\de z$ by \citeN{2010ApJ...721.1448S} 
instead of the previous data by \citeN{2010ApJ...718..392P}. The new data set
includes observations over a wide redshift range
($0.36 < z < 6$) and is well suited for studying the
evolution of reionization.

The likelihood function used in our calculations is given by
\be
L \propto \exp (-{\cal L})
\e
where ${\cal L}$ is the negative of the log-likelihood and estimated 
using the relation
\be
{\cal L} = \f{1}{2}\sum_{\alpha=1}^{N_{\rm obs}} \left[
\f{{\cal J}_{\alpha}^{\rm obs} - {\cal J}_{\alpha}^{\rm th}}{\sigma_{\alpha}}
\right]^2 + {\cal L'}
\e
where  ${\cal J}_{\alpha}$ represents the set of $N_{\rm obs}$ observational data 
points related to photoionization rate and 
distribution of Lyman-limit systems, i.e., ${\cal J}_{\alpha} = \{\log(\Gamma_{\rm PI}), 
\de N_{\rm LL}/\de z\}$, $\sigma_{\alpha}$ are the corresponding 
observational error-bars and $\cal L'$ is negative of WMAP7 or PLANCK log-likelihood function 
for $C_{l}^{\rm TT}$, $C_{l}^{\rm TE}$ and $C_{l}^{\rm EE}$ up to $l=2000$. 
We constrain the free parameters by maximizing the 
likelihood function with a prior that reionization should be completed by $z = 5.8$, 
otherwise it will not match Ly$\alpha$ and Ly$\beta$ forest transmitted flux data.

In this work, we calculate likelihoods using the code
described in Paper I which is essentially based on the publicly available 
COSMOMC\footnote{http://cosmologist.info/cosmomc/} \cite{2002PhRvD..66j3511L} code.
Besides this, throughout we work in a flat cold dark matter model with a cosmological 
constant ($\Lambda$CDM) cosmology with the cosmological parameters given by the current WMAP7 
(based on RECFAST 1.5 \cite{1999ApJ...523L...1S,2000ApJS..128..407S,2008MNRAS.386.1023W} 
and version 4.1 of the WMAP likelihood) best-fit values:
$\Omega_{m}=\Omega_{\rm DM}+\Omega_{b}=0.27$, $\Omega_{\Lambda}=1-\Omega_{m}$, $\Omega_{b}h^2=0.02249$, $h = 0.704$ 
and $\de n_s/\de \ln k =0$ \cite{arXiv:1001.4635v2}. 
Note that, here in all cases, 
$\tau_{\rm el}$ is a derived parameter and the error on obtaining this quantity is slightly 
underestimated because of neglecting the degeneracies between $\tau_{\rm el}$ and other 
cosmological parameters.

\subsection{Brief theory of PCA}
\label{sec:pca_theory}

In this section, we outline the principal component method and introduce the notation 
that we will use throughout the paper. As has been described in Paper I, the
principal components filter out components of the model which are
most sensitive to the data. In order to determine the 
principal components of $N_{\rm ion}(z)$, we consider the data 
for photionization rate $\Gamma_{\rm PI}$, the redshift
distribution of Lyman-limit systems $\de N_{\rm LL}/\de z$ and the
large-scale E-mode polarization angular power spectrum
$C_l^{\rm EE}$ ($l \leq 23$).

We represent the unknown function $N_{\rm ion}(z)$ by a set of $n_{\rm bin}$ discrete
free parameters with the bin width 
\begin{equation}
 \Delta z=\frac{z_{\rm max}-z_{\rm min}}{n_{\rm bin}-1}.
\end{equation}
We have taken a redshift range $[z_{\rm min}:z_{\rm max}] = [0:30]$ and  
$\Delta z = 0.2$ (i.e. $n_{\rm bin} = 151$). Then we construct the Fisher matrix  
\begin{equation}
 F_{ij}=\sum_{\alpha=1}^{n_{\rm obs}}\frac{1}{\sigma_{\alpha}^2}
\frac{\partial {\cal G}_{\alpha}^{\rm th}}{\partial N_{\rm ion}^{\rm fid}(z_i)}
\frac{\partial {\cal G}_{\alpha}^{\rm th}}{\partial N_{\rm ion}^{\rm fid}(z_j)},
\label{eq:fisher_matrix}
\end{equation}
where 
${\cal G}_{\alpha},~\alpha=1,2,\ldots,n_{\rm obs}$ represent the observational
data points (which in our case is given by ${\cal G}_{\alpha} = \{\log(\Gamma_{\rm PI}), \de N_{\rm LL}/\de z,C_l^{\rm EE}\}$),
${\cal G}_{\alpha}^{\rm th}$ is theoretical value of ${\cal G}_{\alpha}$ 
and $N_{\rm ion}^{\rm fid}$ is the fiducial model which is, in principle, close to the 
underlying ``true'' model. In this work we take the fiducial model $N_{\rm ion}^{\rm fid}$ 
to be the model which matches the $\Gamma_{\rm PI}$, $\de N_{\rm LL}/\de z$ and CMB data 
points up to an acceptable accuracy and also which is characterized by a higher 
$N_{\rm ion}$ at higher redshifts. The match with the data for our fiducial model is similar 
to Figure 2 of \citeN{2009arXiv0904.4596C} and Figure 1 of Paper I. 

Once the Fisher matrix is constructed, we can determine its eigenvalues and 
corresponding eigenvectors. Because of the orthonormality and completeness of 
the eigenfunctions, we can expand the deviation of $N_{\rm ion}$ from its fiducial 
model, $\delta N_{i}=N_{\rm ion}(z_i)-N_{\rm ion}^{\rm fid}(z_i)$, as 
\begin{equation}
 \delta N_i=\sum_{k=1}^{n_{\rm bin}} m_kS_{k}(z_i)
\end{equation}
where $S_k(z_i)$ are the principal components of $N_{\rm ion}(z_i)$ and 
$m_k$ are the expansion coefficients. The advantage is that, unlike $N_{\rm ion}(z_i)$, 
the coefficients $m_k$ are uncorrelated with variances.

In realistic situations, there will be other free parameters (apart from
$m_k$ or $\delta N_i$) in the model. Let there be $n_{\rm ext}$ number of extra parameters 
other than $m_k$; this means that we are now dealing with
a total of $n_{\rm tot} = n_{\rm bin} + n_{\rm ext}$ parameters.
In this case, we can still form the Fisher matrix of 
$n_{\rm tot} \times n_{\rm tot}$ dimensions which can be written as
\be
{\cal F} = \left(\begin{array}{cc}
{\matrixsymbol F} & {\matrixsymbol B}\\
{\matrixsymbol B}^T & {\matrixsymbol F'}
\end{array}\right)
\e
where ${\matrixsymbol F}$ is the $n_{\rm bin} \times n_{\rm bin}$-dimensional 
Fisher matrix for the $\delta N_i$, ${\matrixsymbol F'}$ is the
$n_{\rm ext} \times n_{\rm ext}$-dimensional 
Fisher matrix for the other parameters and 
${\matrixsymbol B}$ is a $n_{\rm bin} \times n_{\rm ext}$-dimensional
matrix containing the cross-terms.
One can then invert the above ${\cal F}$ to obtain the corresponding Hessian
matrix ${\cal T} = {\cal F}^{-1}$. Following that, one simply retains 
the sub-block ${\matrixsymbol T}$ corresponding to $\delta N_i$
whose principal components will be ``orthogonalized'' to
the effect of the other parameters. The resulting ``degraded'' sub-block
will be \cite{1992nrfa.book.....P}
\be
{\matrixsymbol \tilde{F}} = {\matrixsymbol T}^{-1}
= {\matrixsymbol F} - {\matrixsymbol B} {\matrixsymbol F'}^{-1} {\matrixsymbol B}^T
\e

In this work we need to use the above formalism to marginalize over the normalization 
of the mean free path $\lambda_0$, cosmological parameters $n_s$ and $\sigma_8$. 
So, in this case, $n_{\rm ext}=3$. 

It can be shown that the largest eigenvalues correspond to minimum variance and 
vice versa. Hence, most of the information relevant for the observed data points is 
contained in the first few modes with larger eigenvalues. We can then reconstruct 
the function $\delta N_{i}$ using only the first $M\leq n_{\rm bin}$ modes. 
So, the important step in this analysis is to decide on how many modes $M$ to be used. 
If we include all the $n_{\rm bin}$ modes, then no information is thrown away, but 
the errors in the recovered quantities would be very large due to the presence of very 
small eigenvalues. On the other hand reducing $M$ can reduce the error but it may introduce 
large biases in the recovered quantities. 

One possible approach is to use the trial-and-error method to fix $M$, i.e. assume an 
underlying model which is different from the fiducial model but matches the current 
data sets quite accurately
and study its recovery using only first few modes. We refer the reader to our earlier 
paper for a detailed discussion about this approach. A slightly more formal approach is 
to estimate $M$ by minimizing the quantity Risk which is essentially
the sum of the bias contribution which arises from neglecting the higher 
order terms, and the error (given by Cramer-Rao bound) arising 
because of higher order terms being included. 
We have checked that the quantity Risk 
has a clear minimum at $M = 8$ for our present case.  

However, both methods described above, involve the assumption of an ``underlying 
model'', hence the determination of $M$ using this method would be model-dependent. 
An alternate prescription is to use Akaike information criterion \cite{2007MNRAS.377L..74L}
\begin{equation}
  {\rm AIC}=\chi_{\rm min}^2+2M
\end{equation}
where smaller values are assumed to imply a more favored model.
Similarly, one can also use the Bayesian information criterion defined by
${\rm BIC}=\chi_{\rm min}^2+M\ln n_{\rm obs}$.
 The utility of these criteria over the Risk is that they are computed without 
knowing the underlying solution \cite{2010PhRvL.104u1301C}. 
The results using BIC typical 
give smooth reconstructions by underestimating the errors. 
The AIC, on the other hand, renders more featured reconstructions at the expense of 
large errors. However, as $n_{\rm obs}$ is fixed for our current analysis, the minimum 
value of AIC corresponds to the minimum of BIC, hence we simply carry out our 
analysis with only AIC. Note that there is no reason to select one 
particular reconstruction, the minimum of AIC can be accompanied by an 
increased chance of getting the reconstructed parameters wrong. 
According to \citeN{2010PhRvL.104u1301C}, 
one successful strategy 
is to select different $M$ which are near the minimum value of AIC and amalgamate 
them equally at the Monte Carlo stage when we compute the errors. In this way, we can reduce the 
inherent bias which exists in any particular choice of $M$. 
We have examined that, in our case, 
the family of different $M$ reconstructions, starting from $M=2$, which satisfy 
\begin{equation}
 {\rm AIC}<{\rm AIC_{\rm min}}+\kappa
\label{eq:aic_criterion}
\end{equation}
where $\kappa=10$ (which corresponds to $M=8$), produces very solid results. 
For alternative data sets, the value of $\kappa$ can be adjusted. The choice of 
this parameter must be treated as a prior. The importance of using the AIC
is that the analysis now becomes non-parametric. The method has been
successfully used in reconstructing the dark energy
equation of state using SN-Ia observations \cite{2010PhRvL.104u1301C}. 

\section{Results}
\label{sec:results}

\subsection{The principal components of $N_{\rm ion}(z)$}

\begin{figure}

  \includegraphics[width=2.6in,height=2.8in, angle=270]{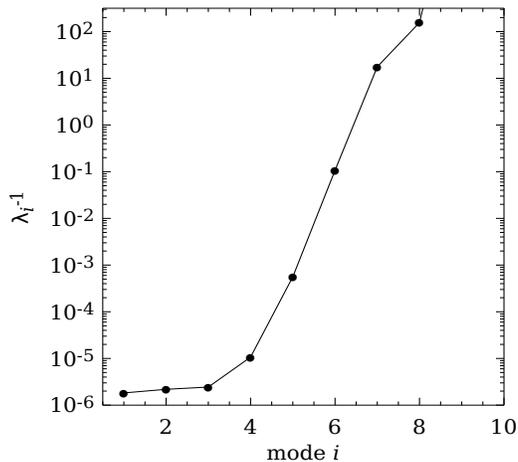} 
  
  \caption{
  The inverse of eigenvalues of the Fisher matrix $F_{ij}$ which essentially measures the variance on the corresponding coefficient.}
\label{fig:eigenval}
\end{figure}

\begin{figure*}
 
  \includegraphics[height=0.6\textwidth, angle=270]{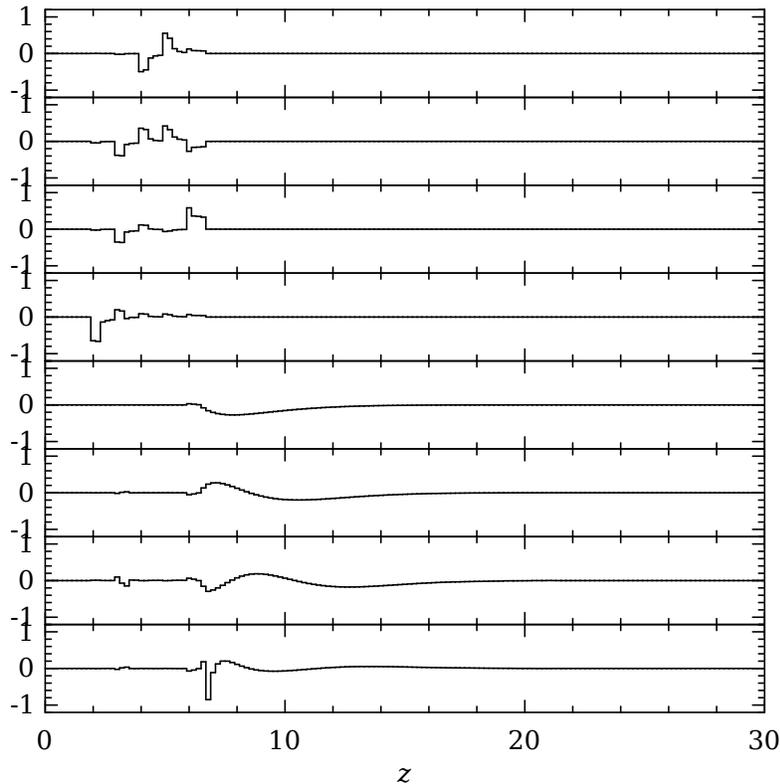}
  \caption{The first 8 eigenmodes of the Fisher matrix.}
\label{fig:eigenvec}
\end{figure*}

The properties of the Fisher matrix $F_{ij}$,
obtained using equation (\ref{eq:fisher_matrix}), were discussed in
detail in Paper I and they remain essentially the same. 
After diagonalizing $F_{ij}$, we obtain its eigenvalues and the corresponding 
eigenmodes. In Figure \ref{fig:eigenval}, 
we show the inverse of the eigenvalues i.e., the variances of the corresponding modes. 
We have verified that the first 5 eigenvalues here are almost the same as those we got in 
our previous work. Interestingly, we get here few more eigenvalues which have considerably  
high values and hence they can not be ignored. 
However, one can see that 7 and 8 modes contain 
less useful information than the first 6 modes. But we have to check first whether
we can simply neglect them or not, because neglecting 7 and 8 modes may
introduce large biases in the recovered quantities. For that, we have used more than one 
method to fix $M$(as described in the earlier section) and each method 
suggests that we should keep upto $M=8$ modes in our analysis unlike the case for Paper I, 
where we got the optimum value of $M$ is five.
This is because the 
$C_l$'s contain somewhat more information than what is contained within
a single data point $\tau_{\rm el}$.
This fact can be noted from the plot of the first 8 eigenmodes (i.e., those
which have the lowest variances) plotted in Figure \ref{fig:eigenvec}. 
The first five modes are similar to what was obtained in Paper I.
However, the modes 6 to 8 in Paper I did not contain any information, while
in this case they show the sensitivity of $N_{\rm ion}(z)$ on different
angular scales $l$.
We find that all the  eigenmodes tend to vanish at $z>15$, which 
is obvious because of $F_{ij}$ being negligible at these redshifts. We can see 
a number of spikes and troughs in the first four modes whose positions correspond to 
the presence of data points for $\Gamma_{\rm PI}$ and $\de N_{\rm LL}/\de z$ 
at $2<z<6$. The last four modes contain the information about the sensitivity 
of $C_l^{\rm EE}$. This sensitivity is maximum 
around $z\approx7-8$ and decreases 
at $z>8$ due to unavailability of free electrons; it also decreases at $z<7$ because of 
the fact that reionization is mostly completed at these redshifts ($x_e \to 1$) and hence changing 
$N_{\rm ion}$ does not affect the value of $C_l^{\rm EE}$ significantly at this 
redshift range. 
The modes ($>8$) with smaller eigenvalues i.e. large variances introduce huge uncertainties 
in the determination of $N_{\rm ion}$ and hence do not contain any meaningful information 
about the reionization history.

\subsection{Markov Chain Monte Carlo Constraints from WMAP7 data}
\label{sec:wmap_likelihood}

The constraints on reionization are obtained by performing a Monte-Carlo Markov Chain (MCMC) 
analysis over the parameter space of the optimum number of PCA amplitudes, $\lambda_0$, $n_s$ 
and $\sigma_8$. Other cosmological parameters are kept fixed to the WMAP7 best-fit values 
(see Section \ref{sec:datasets}). 
To avoid the confusion about the correct choice of number of modes, we perform the MCMC analysis 
for PCA amplitudes taking from $M=2$ to $M=8$, all of which obey the AIC criterion 
(equation \ref{eq:aic_criterion}).  
We then weight each choice of $M$ equally and fold the corresponding errors together 
to reproduce $N_{\rm ion}$ and other related quantities along with their effective errors.   
In order to carry out the analysis, we have developed a code based on the publicly available 
COSMOMC \cite{2002PhRvD..66j3511L}. We run a number of separate chains 
(varying between 5 to 10) until the Gelman and Rubin convergence statistics, $R$, 
corresponding to the ratio of the variance of parameters between chains to the variance 
within each chain, satisfies $R - 1 < 0.01$. Also we have used the convergence diagnostic of 
Raftery \& Lewis to determine how much each chain must be thinned to obtain independent samples. 
Both of these are computed automatically by COSMOMC.

\begin{figure*}
  \includegraphics[height=0.9\textwidth, angle=270]{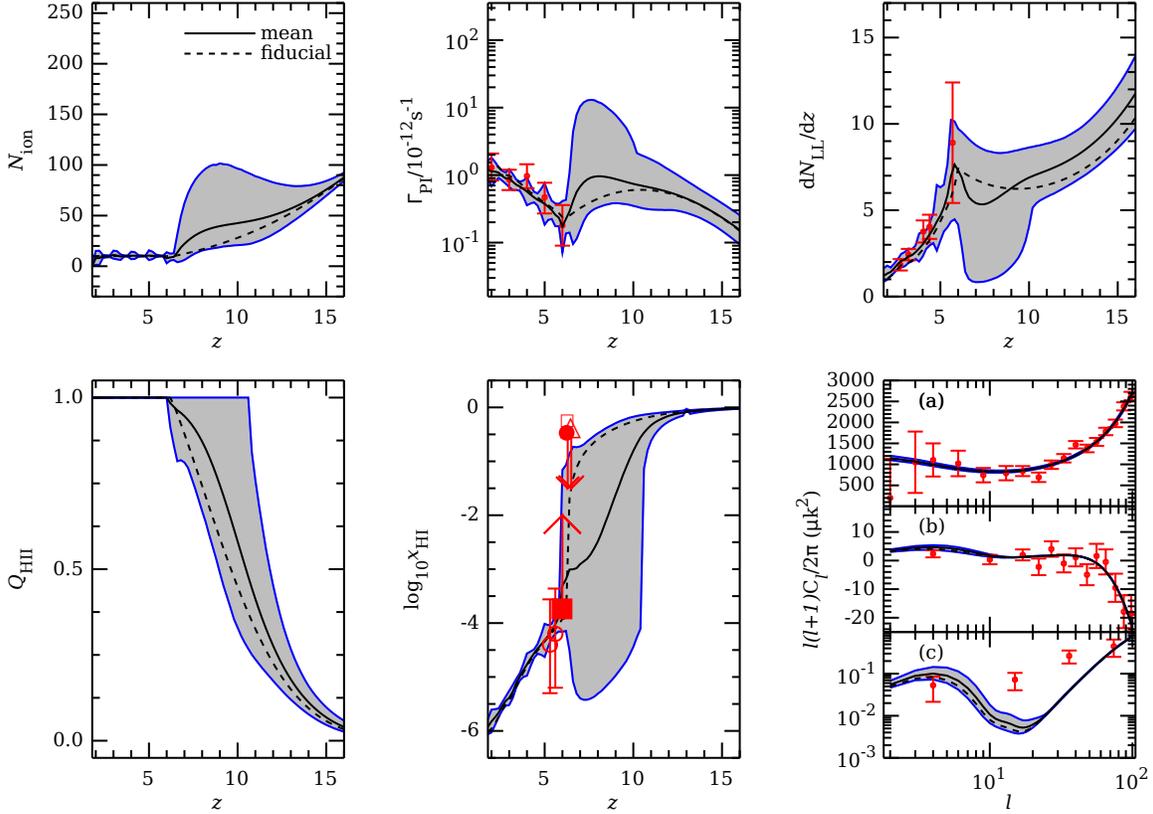}
  \caption{The marginalized posteriori distribution of various quantities related to
  reionization history obtained from the PCA using the AIC criterion with first 8 
  eigenmodes. The solid lines correspond to the model described by mean values of the
  parameters while the shaded regions correspond to 2-$\sigma$ limits.
  The points with error-bars denote the observational data points.
  {\it Top-left:} the evolution of 
  the effective $N_{\rm ion}(z)$;
  {\it Top-middle:} 
  the hydrogen photoionization rate $\Gamma_{\rm PI}(z)$ 
  along with the constraints from Bolton \& Haehnelt (2007);
  {\it Top-right:} 
  the LLS distribution $\de N_{\rm LL}/\de z$ with data points from Songaila \& Cowie (2010); 
  {\it Bottom-left:} 
  the volume filling factor of HII regions $Q_{\rm HII}(z)$;
  {\it Bottom-middle:}
  the global neutral hydrogen fraction $x_{\rm HI}(z)$ with observational
  limits from QSO absorption lines (Fan et al. 2006; filled square), 
  Ly$\alpha$ emitter luminosity function (Kashikawa et al. 2006; open triangle) 
  and GRB spectrum analysis (Totani et al 2006; open square). 
  Also shown the constraints using dark gap statistics on QSO spectra 
  (Gallerani et al 2008a; open circles) and GRB spectra (Gallerani et al. 2008b;
  filled circle);
  {\it Bottom-right:} (a) TT, (b) TE and (c) EE power spectra with the data 
  points from WMAP7 (Larson et al. 2010).
  In addition, we show the properties of the fiducial 
  model (short-dashed lines)
  as described in Section \ref{sec:pca_theory}.
  }
\label{fig:aic_wmapclee}
\end{figure*}

We have shown the evolution of various quantities related to reionization using 
the AIC criterion for $M=2$ to $M=8$ 
in figure \ref{fig:aic_wmapclee}. The solid lines represent 
the mean model while the shaded region correspond to 95\% confidence limits. For
comparison, we have also plotted the fiducial model (short-dashed)
as described in Section \ref{sec:pca_theory}. We find that the fiducial model 
is within the 95\% confidence limits for the whole redshift range.
Note that all the quantities are highly constrained at $z < 6$, 
which is expected as most of the observational information related to reionization
exists only at those redshifts. The errors also decrease at $z > 14$ as there is 
practically no information in the PCA modes and hence all models converge towards 
the fiducial one. The most interesting information regarding reionization is concentrated 
within a redshift range $6 < z < 14$. 

\begin{table}
\begin{tabular}{c|c|c}
Parameters & Mean value & 95\% confidence limits\\
\hline
$\tau_{\rm el}$ & $0.093$ & $[0.080, 0.112]$\\
$z(Q_{\rm HII}=0.5)$ & $10.206$ & $[8.952, 11.814]$\\
$z(Q_{\rm HII}=0.99)$ & $7.791$ & $[5.800, 10.427]$\\
\hline
\end{tabular}
\caption{The marginalized posterior probabilities with 95\% C.L. errors of all the 
derived parameters for the reionization model obtained from the current analysis 
using AIC criterion for WMAP data.   
}
\label{tab:AIC_wmap_likelihood}
\end{table}

It can be seen from the plot of $N_{\rm ion}(z)$ (\textit{top-left} panel of figure 
\ref{fig:aic_wmapclee}) that such quantity must necessarily increase from 
its constant value at $z < 6$ which confirms our findings from Paper I. This rules out the possibility of reionization 
with a single stellar population having non-evolving IMF and/or star-forming efficiency. 
The main difference from our previous results is that the allowed ranges 
in $N_{\rm ion}$ at redshifts $7 < z <12$ has reduced significantly
(earlier, values of $N_{\rm ion}$ as large as 250 were allowed around
$z \approx 7.5$, while the maximum allowed value has been reduced to $\approx 100$
in this work). While some of these constraints arise from the 
observation of Lyman-limit systems at $z \approx 6$, the major
effects arise due to the inclusion of $C_l$'s into the analysis.
This again confirms the fact that $C_l$'s have more
constraining power than $\tau_{\rm el}$ taken as a single point.

The same conclusion can be drawn from the plot of $\Gamma_{\rm PI}(z)$ (\textit{top-middle} panel), where we 
find that the maximum allowed value is $\approx 10^{-11}$ s$^{-1}$.
This is nearly 10 times more stringent than what was allowed in Paper I.
We find that the mean model 
is consistent with the observational data at $z < 6$, as expected. The errors corresponding 
to 95\% confidence limits are also smaller at this epoch. The photoionization rate for the 
fiducial model shows a smooth rise at $z > 6$ reaching a peak around $z \approx 11$; however, the model 
described by the mean values of the parameters shows a much sharper rise and much prominent 
peak around $z \sim 6.5$. The prominent peak-like structure is also present in the plot of 
$\de N_{\rm LL}/\de z$ (\textit{top-right} panel).

From the plot of $Q_{\rm HII}(z)$ (\textit{bottom-left} panel), we see that the growth of $Q_{\rm HII}(z)$
for the mean model is much faster than that of fiducial model at initial stages, though the 
completion of reionization takes place only at $z \approx 6$. One can also find that reionization 
can be completed as early as $z \approx 10.4$ (95\% confidence level). Similarly, $x_{\rm HI}(z)$ 
(\textit{bottom-middle} panel) decreases much faster than the fiducial one at $6<z<12$ and then smoothly 
matches the Ly$\alpha$ forest data. 

Finally, we have shown the values of (a) $C_l^{\rm TT}$, (b) $C_l^{\rm TE}$ and (c) $C_l^{\rm EE}$ 
for the mean model in the \textit{bottom-right} panel of this figure, which is almost the same as the 
fiducial model. So the current WMAP7 EE polarization data alone cannot distinguish between 
the various models of reionization.  One can see that, our mean model includes most of the current 
WMAP7 best-fit CMB data within the error bars, except for a few $C_{l}^{\rm EE}$ data points. Note 
that these discrepant points at $l \gtrsim 15$ cannot be reconciled by any {\it physical} 
reionization model, implying that the spectra contribution might come from some other cosmological 
process, as e.g. gravitational lensing.

The mean values and the 95\% confidence limits on the parameters obtained
from our analysis are shown in the Table \ref{tab:AIC_wmap_likelihood}. 
We have checked that, our fiducial model which is characterized by 
$m_1=m_2=m_3=m_4=m_5=m_6=m_7=m_8=0$ and the best-fit values 
of $\lambda_0$, $n_s$ and $\sigma_8$, is included within
the 95\% confidence limits of those parameters corresponding to our current analyses 
using AIC criterion.
We find that reionization is 50\% complete between redshifts 9.0 -- 11.8 (95\% confidence level), 
while it is almost (99\%) complete between redshifts 5.8 -- 10.4 (95\% confidence level). 
These values are similar to what was obtained in Paper I.
Note that the lower limit on the redshift of reionization (5.8) is imposed as a prior
on the parameters. Here the mean model for $\tau_{\rm el}$ shows a higher value than the best-fit 
WMAP7 value which is arising from relatively complex reionization histories 
giving non-zero ionized fractions at high redshifts.
The value of $\tau_{\rm el}$ obtained is slightly lower than what we got in our earlier work, 
where we included $\tau_{\rm el}$ as a single data point instead of considering CMB large-scale 
EE polarization data which is because many models with very high $N_{\rm ion}$ are ruled out
in this work. 

We have checked that, if we take any particular choice of $M$, say $M=7$ or 8, our main findings 
are almost the same as the above results, except with the help of AIC criterion, we have reduced 
the inherent bias which is present for that specific choice of $M$ and got a mean model which matches 
the current data sets quite reasonably.  

To summarize, we find that using $C_l^{\rm EE}$ data set instead of $\tau_{\rm el}$, we can get 
a relatively smaller error for $N_{\rm ion}(z)$ (see Figure 7 of \citeNP{2010arXiv1011.2213M}) 
but get a $\tau_{\rm el}$ which is higher than the current WMAP value. So a wide range of 
reionization histories is still allowed by the data we have used. Reionization can be 
quite early or can be gradual and late, depending on the behavior of $N_{\rm ion}(z)$. Hence, 
using these data, it is somewhat difficult to put strong constraints on chemical feedback 
and/or the evolution of star-forming efficiencies and/or escape fractions. 

\subsection{Markov Chain Monte Carlo Constraints from simulated PLANCK forecast data}
\label{sec:planck_likelihood}

Given that the current data allow a large range of reionization models, it is
worthwhile computing the level of constraints expected from future
large-scale polarization measurements like those obtained from PLANCK.
To forecast the errors for parameters related to the reionization history, 
we first generate the 
simulated PLANCK data of CMB power spectra for our fiducial model up to $l\leq2000$ using the 
exact full-sky likelihood function at PLANCK-like sensitivity 
\cite{arXiv:astro-ph/0606227,2010PhRvD..82l3504G}. 
We assume that beam uncertainties are small and that uncertainties due to 
foreground removal are smaller than statistical errors.
More sensitive 
observations will also require an exact analysis of 
non-Gaussian likelihood function, here 
for simplicity we assume isotropic Gaussian noise and neglect non-Gaussianity of the 
full sky \cite{2005PhRvD..71h3008L} and try to see 
what we can learn about the global reionization history 
from PLANCK-like sensitivity. We then repeat the MCMC analysis over the same parameter 
space of Section \ref{sec:wmap_likelihood} using this simulated data. Like the previous 
case, here we have also varied the number of modes included in the analysis from two to 
eight using the AIC criterion in order to study the effect of truncating the PCA expansion 
for the recovery of various quantities related to reionization.

\begin{table}
\begin{tabular}{c|c|c}
Parameters & \multicolumn{2}{c}{2-$\sigma$ errors}\\
& WMAP7 & PLANCK (forecast)\\
\hline
\hline
$\tau_{\rm el}$ & $0.032$ & $0.009$\\
$z(Q_{\rm HII}=0.5)$ & $2.862$ & $1.117$ \\
$z(Q_{\rm HII}=0.99)$ & $4.627$ & $3.013$ \\
\hline
\end{tabular}
\caption{The 95\% C.L. errors of derived parameters 
  for the reionization model obtained from the current 
  analyses using AIC criterion for WMAP7  and simulated PLANCK 
  data.}
\label{tab:comparison_likelihood}
\end{table}

In the Table \ref{tab:comparison_likelihood}, we have shown the comparison of 
the 2-$\sigma$ errors on the derived parameters obtained  for currently 
available WMAP7 data and the same for forecasts from simulated PLANCK data. 
It is clear that the uncertainties on all the parameters related to
reionization would be reduced considerably. 
In particular, we find that we should be able to constrain the
redshift range at which reionization was 99\% (50\%) completed to about 
3 (1). This is clearly a significant improvement
over what can be achieved through current data sets.

\begin{figure*}
  \includegraphics[height=0.9\textwidth, angle=270]{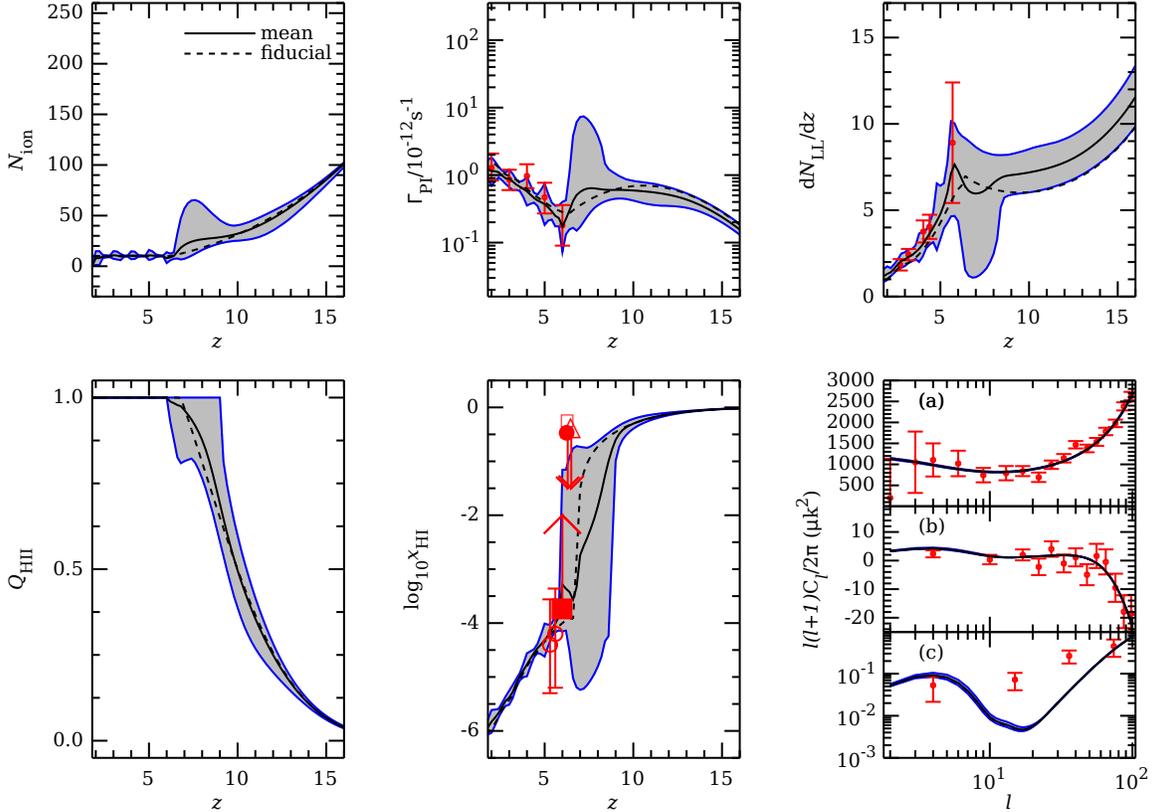}
  \caption{Same as Figure~\ref{fig:aic_wmapclee} but for planck likelihood}
\label{fig:aic_planckclee}
\end{figure*}

In Figure \ref{fig:aic_planckclee}, we have illustrated the recovery the same quantities 
as mentioned in the earlier section using the AIC criterion taking up to 8 eigenmodes for 
the simulated PLANCK data. For comparison, here also we have plotted the results 
for the fiducial model (short-dashed lines) 
along with the mean results (solid lines) from MCMC analysis with shaded 2-$\sigma$ limits.
We find that our main results are in quite reasonable agreement with those obtained from the WMAP 
data (Section \ref{sec:wmap_likelihood}), except that 
all the 2-$\sigma$ (95 \%) limits are reduced remarkably for all redshift range.

We thus find that
we can constrain the global reionization history quite better using the 
PLANCK forecast data sets, especially 
the $2-\sigma$ limits for $Q_{\rm HII}$ reduces significantly for this case.  
However there is no room to substantially improve the constraints 
using large-scale \textit{E}-modes for WMAP7 data sets and one still has to
rely on other types of data for understanding reionization.

\section{Discussion and Summary}
\label{sec:summary}

Based on the work of \citeN{2010arXiv1011.2213M} on principal component analysis of 
reionization model, we have studied constraints on reionization history 
using non-parametric methods. To model the unknown function $N_{\rm ion}(z)$, we have applied  
the principal component method using three different sets of data points - 
the photoionization rate $\Gamma_{\rm PI}(z)$, the LLS distribution $\de N_{\rm LL}/\de z$ 
and current WMAP data for $C_l^{\rm EE}$ for $l \leq 23$.
Following that, we have 
obtained constraints on the reionization history 
using MCMC techniques. We have also used the Akaike information criteria 
  (AIC) to extract the underlying information about the PCA model and reduce the intrinsic 
  bias present in any particular choice of fiducial model. 
We have applied our method to the currently available WMAP7 data as well 
as the simulated PLANCK data to forecast future errors on reionization.

Our main findings can be summarized as follows - 
\begin{enumerate}

 \item We have found that the information about $N_{\rm ion}(z)$ or equivalently the 
  star formation and/or chemical feedback lies in the first eight eigenmodes of the 
  Fisher information matrix distributed over the range $2<z<14$. Using the higher modes 
  costs higher errors. 
  
 \item The angular power spectra $C_l$ of CMB observations contain more information
 than treating $\tau_{\rm el}$ as a single data point. This is obvious from the
 analysis of the Fisher matrix and results in (slightly) more stringent
 constraints on $N_{\rm ion}(z)$ and $\Gamma_{\rm PI}(z)$.

 \item The constraints at $z<6$ are relatively tight because of the QSO absorption line
 data. On the other hand, a wide range of histories at $z> 6$ is allowed by the data.
 Interestingly, it is not possible to match the available data related to reionization with a 
  constant $N_{\rm ion}(z)$ over the whole redshift range, it must increase at $z > 6$ 
  from its constant value at lower redshifts. 

\item With currently available data from WMAP7, we constrain 
$0.080 < \tau_{\rm el} < 0.112$ (95\% CL) and also
conclude that reionization is 50\% complete between 
$9.0 < z(Q_{\rm HII} = 0.5) < 11.8$ (95\% CL) and is 
99\% complete between $5.8 < z(Q_{\rm HII} = 0.99) < 10.4$ (95\% CL).

\item With the forthcoming 
PLANCK data on large-scale polarization (ignoring effect of foregrounds), 
the $z > 6$ constraints will be improved considerably, e.g., 
the $2-\sigma$ error on $\tau_{\rm el}$ will be reduced 
to 0.009 and the uncertainties on $z(Q_{\rm HII} = 0.5)$ and 
$z(Q_{\rm HII} = 0.99)$ would be $\sim 1$ and 3 (95\% CL), respectively.
The errors could be somewhat larger if the effect of foregrounds
are incorporated into the analysis. 
For more stringent constraints on reionization at $z > 6$, one has to rely on data
sets other than CMB.

\end{enumerate}

Finally, we try to indicate the data sets (other than CMB) which can possibly be used 
to better the constraints on reionization. Since most of the information
on reionization at $z < 6$ come from QSO absorption lines, it is natural to
expect more constraints from such observations at $z > 6$. In addition, spectra
of GRBs, which are being observed at much higher redshifts 
\cite{2009Natur.461.1258S,2009Natur.461.1254T,2011arXiv1105.4915C} 
could also provide additional constraints. The difficulty is that the transmission
regions (which are the sources for most of the information) are almost non-existent
at high-$z$ spectra, thus making the analysis more difficult.
Additional constraints on $x_{\rm HI}$ at high redshifts are expected from
Ly$\alpha$ emitters 
\cite{2005PASJ...57..165T,2006ApJ...648....7K,2006Natur.443..186I,2011ApJ...730L..35V,2010Natur.467..940L}, 
however they too are affected highly by systematics. On 
the positive side, we feel that even a relatively weak constraint on 
$x_{\rm HI}$ at $z \sim 7-10$ could be crucial in ruling out a subset of
reionization models as the value of $N_{\rm ion}(z)$ is most uncertain
at these redshifts.

We also now have observations of Lyman-break galaxies till $z \sim 10$ 
\cite{2007ApJ...670..928B,arXiv:1006.4360v3,2011Natur.469..504B}. The
luminosity function of such galaxies would be helpful in constraining 
properties of the galaxies like the IMF and/or the star-forming efficiency.
Unfortunately, that would still leave out the escape fraction
of ionizing photons, which remain an uncertain parameter till date.

Other indirect observations that could help in constraining reionization
are the temperature measurements at $z < 6$ 
\cite{2000MNRAS.318..817S,2000ApJ...534...41R,2001ApJ...562...52M,2001ApJ...557..519Z,2009ApJ...706L.164C}. 
The temperature evolution
can retain memory of how and when the IGM was reionized and thus
could provide additional constraints on reionization. 
Whatever be the case, the principal component method described 
in this paper, could be a promising tool for extracting the information from the future 
data sets in a model-independent manner. 

\section*{Acknowledgements}

Computational work for this study was carried out at the cluster computing facility
in the Harish-Chandra Research Institute\footnote{http://cluster.hri.res.in/index.html}.
TRC would like to thank the Regular Associateship Programms of the ICTP (Trieste)
during which the work was completed.

\bibliography{mnrasmnemonic,IGM-ADS}
\bibliographystyle{mnras}

\end{document}